\documentclass[namedreferences, linksfromyear]{solarphysics}
\usepackage[optionalrh,solaromanenum,natbib,linksfromyear]{spr-sola-addons} % For Solar Physics 
\usepackage{graphicx}                    % For eps figures, newer & more powerfull
\usepackage[usenames, dvipsnames]{color}                       % For color text: \color command
\usepackage{url}                         % For breaking URLs easily trough lines

\usepackage[pdfborder={0,0,0},urlcolor=blue,breaklinks]{hyperref}
\begin{document}

\begin{article}

\begin{opening}

\title{Buoyant Magnetic Loops Generated by Global Convective Dynamo Action}

%%%%%%%%%%%%%%%%%%%%%%%%%%%%%%%%%%%%%%%%%%%%%%%%%%%
%% Authors Names
%
\author{Nicholas~J.~\surname{Nelson}$^{1}$\sep
        Benjamin~P.~\surname{Brown}$^{2,3}$\sep
        A.~Sacha~\surname{Brun}$^{4}$\sep
        Mark~S.~\surname{Miesch}$^{5}$\sep
        Juri~\surname{Toomre}$^{1}$}

%%%%%%%%%%%%%%%%%%%%%%%%%%%%%%%%%%%%%%%%%%%%%%%%%%%
%% Runningheads
%
\runningauthor{{\it N. J. Nelson et al.}}
\runningtitle{Buoyant Magnetic Loops}

%%%%%%%%%%%%%%%%%%%%%%%%%%%%%%%%%%%%%%%%%%%%%%%%%%%
%% Affilations 
%
  \institute{$^{1}$ JILA and Dept. Astrophysical \& Planetary Sciences, University of Colorado, Boulder, CO 80309-0440 USA email: \url{nnelson@lcd.colorado.edu} \\ 
             $^{2}$ Dept. Astronomy, University of Wisconsin, Madison, WI 53706-1582 USA \\
             $^{3}$ Center for Magnetic Self Organization in Laboratory and Astrophysical Plasmas, University of Wisconsin, 1150 University Avenue, Madison, WI 53706 USA\\
             $^{4}$ Laboratoire AIM Paris-Saclay, CEA/Irfu Universit\'e Paris-Diderot CNRS/INSU, 91191 Gif-sur-Yvette, France \\
             $^{5}$ High Altitude Observatory, NCAR, Boulder, CO 80307-3000 USA\\
             }

%%%%%%%%%%%%%%%%%%%%%%%%%%%%%%%%%%%%%%%%%%%%%%%%%%%
%%% Abstract 
\begin{abstract}
Our global 3D simulations of convection and dynamo action in a Sun-like star reveal that persistent wreaths of strong magnetism can be built within the bulk of the convention zone.  Here we examine the characteristics of buoyant magnetic structures that are self-consistently created by dynamo action and turbulent convective motions in a simulation with solar stratification but rotating at three times the current solar rate. These buoyant loops originate within sections of the magnetic wreaths in which turbulent flows amplify the fields to much larger values than is possible through laminar processes. These amplified portions can rise through the convective layer by a combination of magnetic buoyancy and advection by convective giant cells, forming buoyant loops. We measure statistical trends in the polarity, twist, and tilt of these loops. Loops are shown to preferentially arise in longitudinal patches somewhat reminiscent of active longitudes in the Sun, although broader in extent. We show that the strength of the axisymmetric toroidal field is not a good predictor of the production rate for buoyant loops or the amount of magnetic flux in the loops that are produced. 
\end{abstract}

%%%%%%%%%%%%%%%%%%%%%%%%%%%%%%%%%%%%%%%%%%%%%%%%%%%
%% Keywords
%
%\keywords{}

\end{opening}
%-------------------------------------------------

%%%%%%%%%%%%%%%%%%%%%%%%%%%%%%%%%%%%%%%%%%%%%%%%%%%
%% Sections
%

 \section{Flux Emergence and Convective Dynamos}
 
Convective dynamo action in the interior of the Sun is the source of the magnetism that creates sunspots and drives space weather.  Such magnetism is not limited to the Sun, as magnetic activity is observed to be ubiquitous among Sun-like stars.  To understand the origin of sunspots and starspots, the processes which generate magnetic structures and then transport them through the convection zone to the surface must be explored.  Here we present the results of a global numerical simulation, called case S3, which self-consistently generates wreaths of strong magnetic field by dynamo action within the convective zone.  Case S3 models the convection zone of a Sun-like star nominally rotating at three times the current solar rate, or $3 \, \Omega_\odot$.  The wreaths reverse polarity in a cyclic fashion, yielding cycles of magnetic activity. Portions of these wreaths form buoyant magnetic structures, or loops, which rise through our convective envelope.  Initial results on the behavior of a small number of these loops were reported by \cite{Nelson2011}.

Here we discuss the properties of a much larger number of loops in order to get a statistical description of their properties.  We find coherent magnetic structures with a variety of topologies, latitudinal tilts, twists, and total fluxes.  Additionally, we observe only a weak correlation between the unsigned magnetic flux in a buoyant loop and the axisymmetric toroidal magnetic field at that latitude and time, indicating that the generation mechanism for these loops relies on local, coherent toroidal-field structures amplified by turbulent intermittency rather than large-scale instabilities of axisymmetric fields.  We also find evidence for longitudinal intervals which preferentially produce buoyant loops, hinting at a possible origin for active longitudes for sunspots \citep{Henney2002}, although our intervals are quite broad.

Our work builds upon a series of simulations that consider the dynamics within the deep convective envelopes of young suns that rotate faster than our current Sun. Strong differential rotation was found in hydrodynamic simulations involving a range of rotation rates up to $10 \, \Omega_\odot$ \citep{Brown2008}, including prominent longitudinal modulation in the strength of the convection at low latitudes. Turning to dynamo action achieved in a MHD simulation in such stars at $3 \, \Omega_\odot$, \cite{Brown2010} reported that the convection can build global-scale magnetic fields that appear as wreaths of toroidal magnetic field of opposite polarity in each hemisphere. These striking magnetic structures persist for long intervals despite being embedded within a turbulent convective layer.  At a faster rotation rate of $5 \, \Omega_\odot$, self-consistently generated magnetic wreaths at low latitudes underwent reversals in global magnetic polarity and cycles of magnetic activity \citep{Brown2011}. These cyclic reversals can also be achieved at lower rotation rates if the diffusion is decreased, as the reversals can only occur when resistive diffusion is not able to prevent reversals in the axisymmetric poloidal fields \citep{Nelson2012}.  As diffusion is decreased, the level of turbulent intermittency rises, leading to coherent magnetic structures that can become buoyant\citep{Nelson2011}.

Although the simulation discussed here describes Sun-like stars which nominally rotate faster than the current Sun, the dynamo action realized here may not be only confined to rapidly rotating stars.  The most important non-dimensional parameter for the generation of magnetic wreaths is the Rossby number (the ratio of convective vorticity to twice the frame rotation rate) which is small in both the Sun and our simulation here.  While no simulation can achieve solar-like values of all relevant parameters, the ability to self-consistently capture a wide range of dynamics, including the buoyant transport of magnetic structures through the convective layer, provides us with a unique tool for exploring dynamo action in a solar-like context. Thus our work may be broadly applicable also to processes occurring in the solar interior.

\subsection{Magnetism in Many Settings}

Magnetic activity and cycles appears to be characteristic of many Sun-like stars.  The best studied example is clearly the Sun's 22-year magnetic-activity cycle. The interplay of turbulent convection, rotation, and stratification in the solar convection zone creates a cyclic dynamo which drives variations in the interior, on the surface, and throughout the Sun's extended atmosphere \citep{Charbonneau2010}. Yet the Sun is not alone in its magnetic variability. Solar-type stars generate magnetism almost without exception.  Observations reveal a clear correlation between rotation and magnetic activity, as inferred from proxies such as X-ray and chromospheric emission \citep{Saar1999, Pizzolato2003, Wright2011}. However, superimposed on this trend are considerable variations in the presence and the period of magnetic-activity cycles.  There have been a number of attempts to monitor the magnetic-activity cycles of other stars using solar-calibrated proxies for magnetic activity \citep[{\it e.g.} ][]{Baliunas1995, Hempelmann1996, Olah2009}. Improved observational techniques include spot-tracking from {\it Kepler} photometry \citep{Meibom2011, Llama2012} and Zeeman-Doppler imaging \citep{Petit2008, Gaulme2010, Morgenthaler2012}. These are beginning to provide assessments of the size, frequency, and magnetic flux of starspots and the topology and spatial variability of photospheric magnetic fields. 

\subsection{Theoretical Approaches to Solar and Stellar Dynamos}

The solar dynamo is nonlinear, three-dimensional, and involves a wide range of scales in both space and time, but the basis for most theoretical explorations of the solar dynamo comes from mean-field theory \citep{Parker1955, Moffatt1978, Krause1980}.  In these models, toroidal field is generated through the $\Omega$-effect as differential rotation shears large-scale poloidal field into a band of toroidal field in each hemisphere.  Poloidal field is created through a nonlinear interaction parameterized by the $\alpha$-effect.  A wide variety of mechanisms for the $\alpha$-effect have been proposed, some of which rely on the rise of buoyant magnetic loops to form active regions. In the Babcock--Leighton model, for example, buoyant transport of toroidal magnetic flux provides the mechanism for the regeneration and reversal of the poloidal magnetic field \citep{Babcock1961, Leighton1964}. In mean-field models, magnetic buoyancy is parameterized, assuming that a constant fraction of magnetic flux escapes or that flux emergence is triggered when mean fields achieve a certain magnitude \citep[see review by][]{Charbonneau2010}.

 There have been two main numerical approaches to the study of dynamo action and the source of active regions.  The first class of models tracks the rise of buoyant magnetic structures that have been inserted into stratified domains and then allowed to rise \citep[{\it e.g.}  ][]{Caligari1995, Fan2008, Jouve2009, Weber2011}, or alternatively use forced shear layers to create magnetic structures, which then rise buoyantly \citep[{\it e.g.} ][]{Cline2003, Vasil2009, Guerrero2011}. The second class of models has focused on global-scale convective-dynamo processes that generate magnetic structures in the deep interior and may produce cycles of magnetic activity \citep[{\it e.g.} ][]{Browning2006, Brown2010, Ghizaru2010, Brown2011, Racine2011}. Recently \cite{Miesch2012} have explored 3D convective-dynamo action with a Babock--Leighton term in order to include flux transport by means of a parameterization of magnetic buoyancy. Our study here belongs to the second class, using convective-dynamo simulations to produce buoyant magnetic loops. The first account of such modeling was reported by \cite{Nelson2011, Nelson2012}.

 \section{Nature of the Simulation}
 
   \begin{figure} 
 \begin{center}
 \centerline{\includegraphics[width=\linewidth]{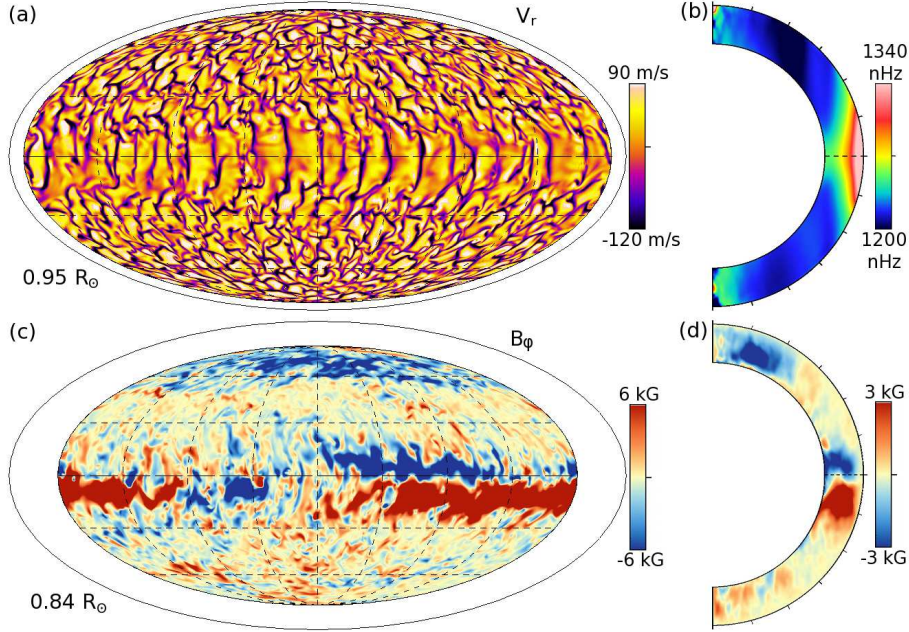}}
 \caption{(a)~Snapshot of radial velocities $v_r$ at time $t _1 = 716$ days in case S3 on a spherical surface at $0.95 R_\odot$ shown in Mollweide projection (Equator at center, lines of constant latitude parallel) in proportional size (outer ellipse represents photosphere). (b)~Rotation rate $\Omega$ averaged in longitude and time. Strong differential rotation is achieved in radius and latitude over the simulated domain. (c)~Companion snapshot of toroidal magnetic field $B_\phi$ at $0.84 R_\odot$, with a strong coherent magnetic wreath in each hemisphere (blue negative, red positive, ranges labeled), with considerable small-scale fields also present. (d)~Azimuthally averaged toroidal magnetic field $\langle B_\phi \rangle$ at the same instant. Low-latitude wreaths are evident in both hemispheres. }
 \label{fig:VrBphi}
\end{center}
 \end{figure}
 
We use the 3D anelastic spherical harmonic (ASH) code to model large-scale convective-dynamo action in the solar convective envelope.  ASH solves the anelastic MHD equations in rotating spherical shells \citep{Clune1999, Brun2004}.  ASH is limited to the deep interior due to the anelastic approximation, which limits us to low Mach-number flows.  Additionally we stay away from the near-surface layers because we cannot resolve the small scales of granulation and super granulation realized near the photosphere.  Our simulation extends from 0.72 $R_\odot$ to 0.965 $R_\odot$, covering a density contrast of about 25 from top to bottom. The details of the numerical scheme used in case S3 are described by \cite{Brown2010}, and the specific parameters are given by \cite{Nelson2012}.  Of special note, in case S3 the Rossby number is 0.581, which is in the same rotationally influenced regime as the giant-cell convection realized in the solar interior \citep{Miesch2005}. Thus the dynamics in case S3 may be broadly applicable to stars like the Sun in which rotational influences on convective motions are significant.
 
 To achieve very low levels of diffusion, we employ a dynamic Smagorinsky subgrid-scale (SGS) model which uses the self-similar behavior in the inertial range of the resolved turbulent cascade to extrapolate the diffusive effects of unresolved scales. In this model the viscosity at each point in the domain is proportional to the magnitude of the strain rate tensor and the constant of proportionality is determined using the resolved flow and an assumption of self-similiar behavior. A detailed description of the dynamic Smagorinsky SGS model is provided in Appendix A of \cite{Nelson2012}. Here we employ constant SGS Prandtl and magnetic Prandtl numbers of 0.25 and 0.5, respectively. In practice this permits a reduction in the average diffusion by about a factor of 50 compared to a simulation with identical resolution and a less complex SGS model, such as in \cite{Brown2011}. This reduction in diffusion is critical not only in enhancing the turbulent intermittency of the magnetic field, but also in permitting the buoyant loops to rise through the convective layer without diffusive reconnection altering their magnetic topology.
 
   \begin{figure} 
 \begin{center}
 \centerline{\includegraphics[width=\linewidth]{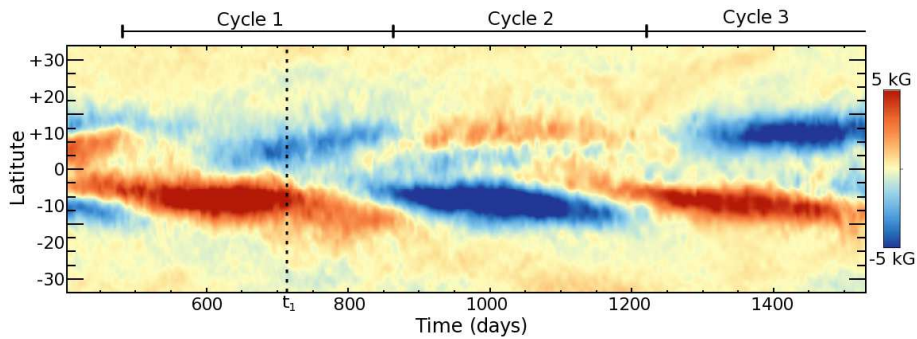}}
 \caption{Evolution in time of the longitudinally averaged toroidal magnetic field $\langle B_\phi \rangle$ at mid-convection zone shown in a time--latitude plot. Three magnetic reversals are realized, each with a period of about 280 days (reversals indicated by hash marks, cycles labeled 1\,--\,3 for convenience). Considerable asymmetry is seen between hemispheres in both the phase and amplitude of the reversals. The time $t_1$ at which snapshots in Figures \ref{fig:VrBphi} and \ref{fig:Wreaths Loops} are sampled is indicated by the dotted line at 716 days.}
 \label{fig:LatTime}
\end{center}
 \end{figure}
 
 Figure~\ref{fig:VrBphi}(a) shows a snapshot of the convective radial velocities $v_r$ in case S3 at a single instant. The convection near the Equator is dominated by convective rolls aligned with the rotation axis, while the higher latitudes have more vortical motions. The rotational influence on the convective motions is key to achieving a pronounced differential rotation \citep{Miesch2006}. Case S3 maintains strong gradients in angular velocity $\Omega$ (Figure~\ref{fig:VrBphi}(b)), which are key to generating the large-scale magnetic wreaths through the $\Omega$-effect. Figures~\ref{fig:VrBphi}(c\,--\,d) show snapshots of the wreaths, both on a spherical surface at mid-convection zone and in their axisymmetric component. The wreaths are dominated by non-axisymmetric fields and thus have a limited longitudinal extent, while clearly still retaining global coherence. 
 
Remarkably, the wreaths are generated and maintained in the bulk of the convective layer without a tachocline of shear. It had been reasonably postulated that coherent, large-scale fields in the convection zone would be shredded by the intense turbulence of the convective motions. However, the convective turbulence evidently does not destroy the wreaths.  In fact, \cite{Nelson2012} showed that while the axisymmetric fields show some decrease in amplitude with increased turbulence, regions of extremely strong fields actually become more common due to increased turbulent intermittency. In regions of particularly strong magnetic fields, the convective motions are diminished by the Lorentz force, resulting in even less convective disruption of the wreaths. 
 
 The dynamic Smagorinsky procedure requires additional computational expense, limiting the time evolution of our simulations.  Case S3 presented here was run for 3.4 million time steps, with an average of 40 seconds of simulated time per step. In total, case S3 covers about four years of simulated time, compared to the rotational period of 9.3 days and the convective over-turning time of about 50 days. Figure~\ref{fig:LatTime} shows the time evolution of the axisymmetric toroidal magnetic field $\langle B_\phi \rangle$ in case S3 over about 1100 days. In this interval there are three reversals of global magnetic polarity. While the true polarity cycle involves two reversals, we term the interval between each reversal an activity cycle in the same way that the Sun's 11-year activity cycles are just about half of the true 22-year polarity cycle. These three activity cycles have durations of about 280 days, although the reversals are not generally synchronized between the two hemispheres. This nonuniform behavior hints at the important role of asymmetries in the flows between the two hemispheres \citep{DeRosa2012}.
 
 \section{Identifying Magnetic Loops}
 
In order to provide a consistent treatment, we define a magnetic loop as a coherent segment of magnetic field that extends from below $0.80 R_\odot$ to above $0.90 R_\odot$ and back down again \citep{Nelson2011}.  Additionally, we require that the buoyant loops have peak magnetic-field strengths greater than 5 kG above $0.90 R_\odot$ at selected samples in time.  To find magnetic loops fitting that description, we have developed a pattern-recognition algorithm that searches the 3D volume of our simulation. The most direct method of finding loops is to look for magnetic-field lines which pass through a region where $\left| B_\phi \right| > 20$ kG below $0.80 R_\odot$, then pass through a region above $0.90 R_\odot$  with $\left| B_\phi \right| > 5$ kG, and then again through a region where $\left| B_\phi \right| > 20$ kG below $0.80 R_\odot$ over less than $50^\circ$ in longitude.  In practice, this can be done much more efficiently by recognizing that the loops start as primarily toroidal magnetic-field structures, but that as they rise into a region of faster rotation the loops are tilted in longitude so that one side of the loop retains a strong component of $B_\phi$ while the other becomes almost totally radial.  Thus we initially identify loop candidates by looking for this pattern of $B_\phi$ and $B_r$.  The loop candidates are then verified using field line tracings.

Case S3 uses 1024 grid points in longitude, 512 in latitude, and 192 in radius for eight evolution variables (velocity $\mathbf{v}$, magnetic field $\mathbf{B}$, entropy $S$, and pressure $P$), thus each snapshot in time requires over 3 GB of data.  We are therefore limited in the number of time steps we can analyze.  For the 278 days of cycle 1 we have run our loop-finding procedure on snapshots of the simulation spaced roughly every four days. In doing so we have identified 131 buoyant loops.  Additionally we have sampled cycle 2 for 20 days and cycle 3 for 40 days with the same four day cadence and found 27 additional loops. We anticipate that we would find many more loops if we carried out a more complete search through cycles 2 and 3.

    \begin{figure} 
 \begin{center}
 \centerline{\includegraphics[width=\linewidth]{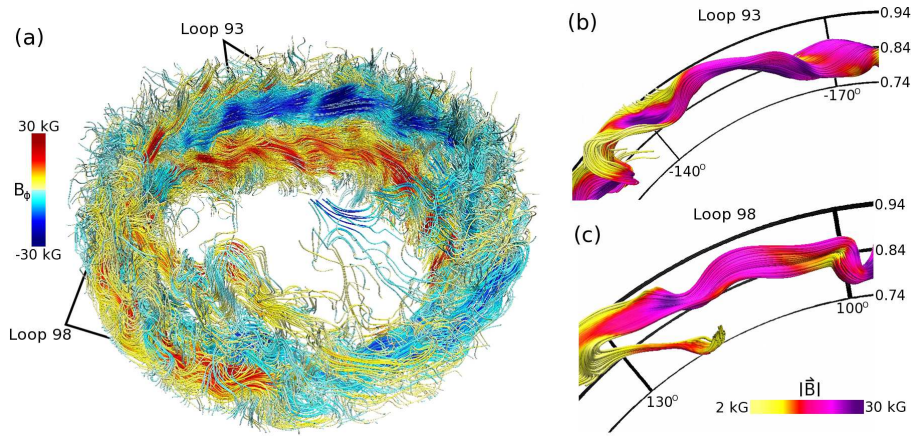}}
 \caption{(a)~Volume-renderings of magnetic-field lines at low latitudes colored by toroidal field $B_\phi$ (red positive, blue negative, amplitudes labeled). Strong magnetic wreaths exist in each hemisphere with considerable modulation in longitude. The location of two sample buoyant loops (labeled loops 93 and 98) are indicated. In this view it is difficult to distinguish the loops from the surrounding magnetic fields. (b\,--\,c) Close-up views of loops 93 and 98 at the same instant with only field lines comprising the buoyant loops rendered for visual clarity. Grid-lines in radius (in units of $R_\odot$) and longitude are provided. Color shows magnitude of magnetic-field strength (yellow weak, purple strong).  Loop 93 is part of the negative polarity wreath in the northern hemisphere, while loop 98 is part of the positive polarity wreath in the southern hemisphere. Time shown corresponds to the snapshots in Figure \ref{fig:VrBphi} and $t_1$ in Figure \ref{fig:LatTime}.}
 \label{fig:Wreaths Loops}
\end{center}
 \end{figure}

For a subset of the 158 loops found in case S3, we have carried out detailed analyses of the dynamics of the rise of 22 of the loops (11 from cycle 1 and 11 more from cycle 3). To do this we have used data with a time resolution of about ten hours, which is sufficient to track loops backward in time from their peak radial position to their origins in the magnetic wreaths.  We find that while the specific evolution of each of these 22 loops varies due to the chaotic nature of the turbulent convection, all 22 loops have significant acceleration due to magnetic buoyancy and are embedded in convective upflows which aid their rise. This agrees with the dynamics of the sample loop studied in detail by \cite{Nelson2011}. While we cannot with certainty say that magnetic buoyancy was a significant factor in the rise of all 158 magnetic loops, we find that for all 22 of the loops studied at high time resolution the average ascent speed due to magnetic buoyancy alone is at least 28\,\% of the total average ascent speed. Thus we assume that magnetic buoyancy is at least an important factor in the rise of these loops.

Figure~\ref{fig:Wreaths Loops} displays the complex nature of the magnetic fields in case S3 with a volume rendering of magnetic-field lines in the convection zone at low latitudes, forming two prominent magnetic wreaths of opposite polarity. We also indicate the location of two buoyant magnetic structures, labeled loops 93 and 98. The simulation continuously exhibits magnetic fields throughout the convection zone, including strong, small-scale magnetic fields, coherent buoyant loops, and large-scale wreaths with global scale organization. Prior studies of magnetic buoyancy typically involved specified buoyant magnetic structures whose rise was studied in a largely unmagnetized domain. In contrast, our convection zone has on average 77\,\% of our simulated volume containing magnetic fields in excess of 1.5 kG, and 21\,\% posses field amplitudes in excess of 5 kG. This makes identification of the buoyant loops difficult. Figures~\ref{fig:Wreaths Loops}(b-c) show close up renderings of only the field lines comprising buoyant loops 93 and 98. We have omitted rendering other field lines in those regions for visual clarity. Magnetic fields in the loops can be quite strong even near the top of our domain, with portions of loop 93 exceeding 25 kG at $0.92 R_\odot$. 
 
 \section{Properties of Rising Loops}

Unlike many previous models of buoyant magnetic transport in which convective turbulence is presumed to play a purely disruptive role, the buoyant loops in our models fall under the turbulence-enhanced magnetic-buoyancy paradigm discussed by \cite{Nelson2012}.  In this model turbulent intermittency plays a key role in the formation of strong, coherent structures that are magnetically buoyant and can be advected by convective upflows. As was shown by \cite{Nelson2011}, these loops rise through a combination of magnetic buoyancy and advection by giant cell convection.  Thus convection plays a key role both in the dynamo which generates the buoyant magnetic fields, and also in the transport of the magnetic loops. Due to the cooperation between convective motions and magnetic buoyancy, the loops are able to rise from below $0.80 R_\odot$ to above $0.90 R_\odot$ in as little as 12 days, as suggested by Figure~\ref{fig:Rise Time}. 

\subsection{Dynamics and Timing of Loop Ascents}

  \begin{figure} 
 \begin{center}
 \centerline{\includegraphics[width=0.95\linewidth]{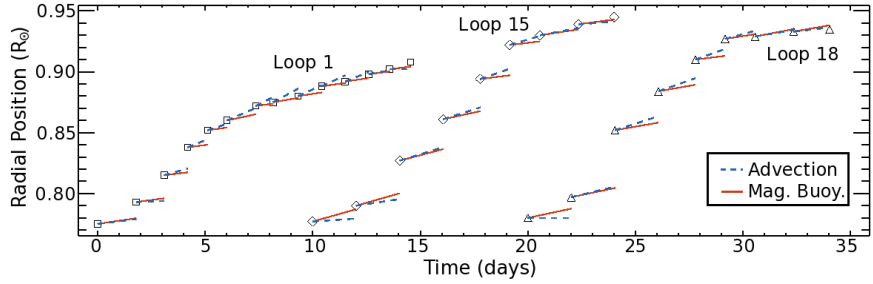}}
 \caption{Location of three buoyant loops (labeled loops 1, 15, and 18) as a function of time as they rise from the core of the toroidal wreaths in the lower convection zone through the simulated domain to their peak radial positions between 0.91 and 0.95 $R_\odot$. Times are given relative to the launch of the loops with offsets for clarity. Also plotted are the mean motions of the loops at each time interval due to magnetic buoyancy (red lines) and advection by the surrounding convective upflows (blue line). Additional motions due to forces such as thermal buoyancy, viscous drag, and magnetic tension are not plotted, and account for what may appear to be missing in this display.}
 \label{fig:Rise Time}
\end{center}
 \end{figure} 

Buoyant loops are born from the much larger and less coherent magnetic wreaths shown in Figures \ref{fig:VrBphi} and \ref{fig:Wreaths Loops}.  The wreaths in case S3 are not axisymmetric structures and are typically coherent over spans of between $90^\circ$ and $270^\circ$ in longitude.  Wreaths exhibit a high degree of magnetic connectivity with the rest of the convection zone, with field lines threading in and out, suggesting rather leaky overall structures.  Wreaths in case S3 generally have average field strengths of between 10 and 15 kG  and are confined in the lower half of the convection zone by magnetic pumping. In the core of the wreaths convective motions can be limited by Lorentz forces to as little as a meter per second.

   \begin{figure} 
 \begin{center}
 \centerline{\includegraphics[width=0.82\linewidth]{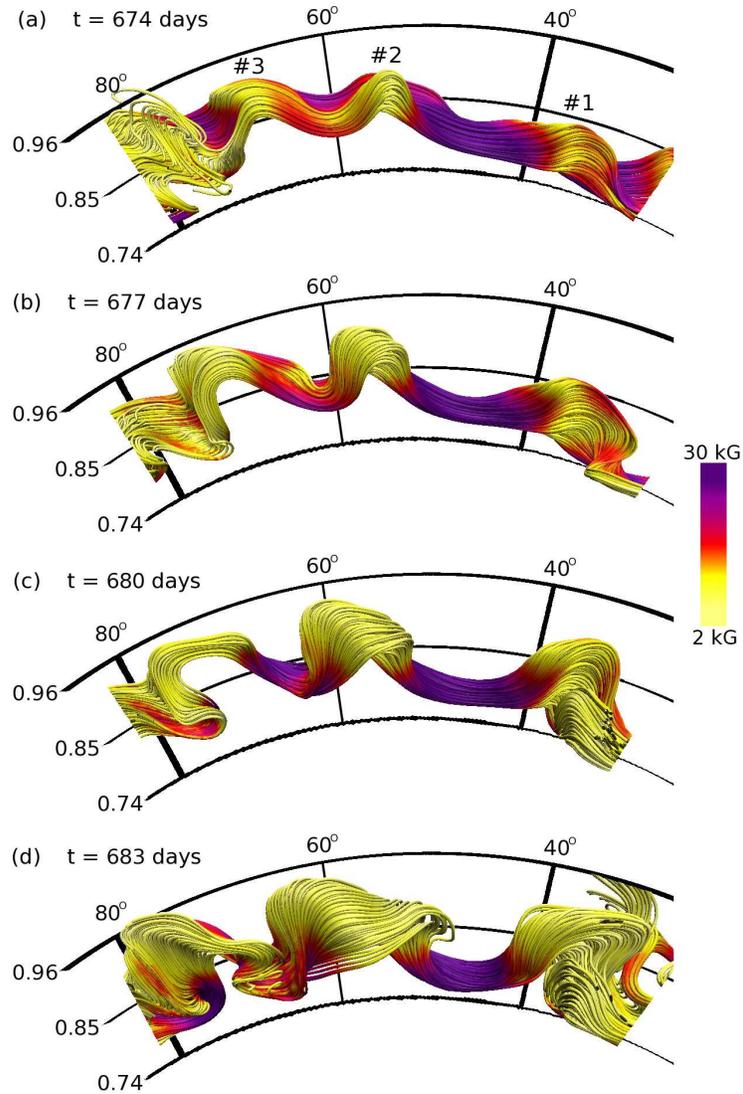}}
 \caption{Sequence of volume rendering of magnetic-field lines which comprise three buoyant loops (labeled loops 1, 2 and 3) as  they rise through the convective layer with three days between each frame (times indicated, progressing downward). Grid lines show radius (in units of $R_\odot$) and longitude. The expansion of each loop is evident here as they rise. Magnetic buoyancy and advection by convective upflows allow the loops to traverse the radial interval shown here in roughly 15 days. Loop 1 is also shown in Figures~\ref{fig:Rise Time} and \ref{fig:Rising Flux}. Loop 3 is also shown in Figure~\ref{fig:3 Views}.}
 \label{fig:Rising Loops}
\end{center}
 \end{figure} 

Portions of these wreaths can be amplified by intermittency in convective turbulence.  Turbulence has been shown to generate strong, coherent structures in a variety of settings \citep{Pope2000}.  In case S3 localized portions of the wreaths are regularly observed to attain field strengths of 40 kG and be highly coherent over as much as $50^\circ$ in longitude.  These magnetic structures with strong fields are able to rise into regions where vigorous convective motions are present. Many structures are seen to emerge from the core of the wreaths only to be pummeled by a convective downflow, disrupted by a region of unusually strong turbulence, or limited by the development of a particularly unfavorable magnetic configuration.  Whether any given magnetic structure becomes a buoyant magnetic loop is therefore not due to the passing of some threshold, but largely a conspiracy of favorable events.

Figure \ref{fig:Rise Time} shows the radial location of the top of three different loops as they rise from roughly $0.77 R_\odot$ to above $0.90 R_\odot$.  Also plotted are the contributions to the radially outward motion due to magnetic buoyancy and advection by convective upflows. The buoyant acceleration due to magnetism is deduced by comparing the density in the region within the loop and the density of the surrounding convective plume. We do this to separate magnetic- and thermal-buoyancy effects. Each of these three loops starts in a region where convective motions are largely suppressed by Lorentz forces due to the very strong magnetic fields in the cores of the magnetic wreaths.  As they begin to rise, the magnetic energy at the core of the wreath exceeds the kinetic energy of the flows locally by a factor of 10 to 100. As the loops rise, they enter regions of strong upflows and are advected upwards by the convective giant cells.  Averaged over their entire ascent, magnetic buoyancy drives an average upward speed of about $50 \; \mathrm{m} \; \mathrm{s}^{-1}$ for these three loops, in addition to the surrounding upflows which move at an average of about $80 \; \mathrm{m} \; \mathrm{s}^{-1}$. At their maximum radial extent, the loops are prevented from rising further by our impenetrable upper boundary condition.

Figure \ref{fig:Rising Loops} shows three sample loops (labeled loops 1, 2, and 3) as they rise over 10 days.  The loops remain coherently connected as they rise. Here again all three loops are aided by convective upflows while convective downflows pin the ends of the loops downward. The direction of motion is largely radial with a deflection of as much as $10^\circ$ in latitude toward higher latitudes.  This deflection is largely due to the roughly cylindrical differential rotation contours realized in this simulation. 

Loops expand as they rise through the stratified domain, but less than would be expected for a purely adiabatic rise. Without any diffusion or draining of material along the field lines, the cross-sectional area of the loops should be inversely proportional to the background pressure, leading to expansion by roughly a factor of 20.  Instead loops are seen to expand by a factor of 5. This is consistent with previous studies of buoyant magnetic structures in which expansion of magnetic structures is seen to be inversely proportional to the square root of the change in pressure \citep{Fan2001, Cheung2010} .

The expansion of the loops is slowed by draining flows of higher entropy fluid along magnetic-field lines, which serves to cool the material at the top of the loop.  These divergent flows are too small to be measured in individual loops due to the turbulent background, but when averaged over 158 loops, a mean divergent flow of 47 cm s$^{-1}$ is obtained along the top of the loops.  This compares well with estimates from a simple model (neglecting viscosity and thermal diffusion) which assumes that the draining flows are constant in time and uniform perpendicular to the axis of the loop.

   \begin{figure} 
 \begin{center}
 \centerline{\includegraphics[width=\linewidth]{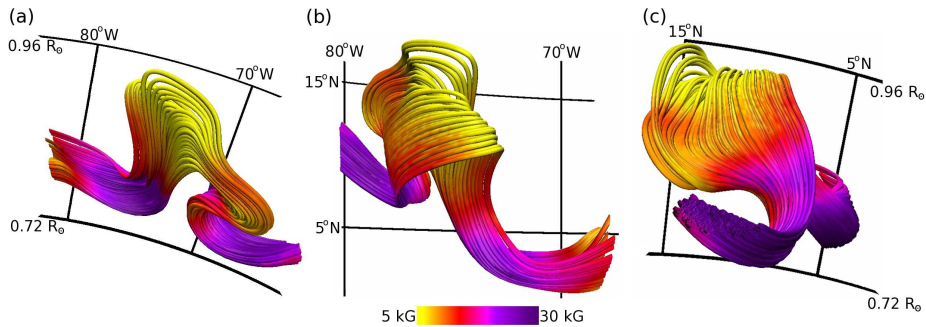}}
 \caption{Three viewpoints the same volume-rendering of magnetic field lines in Loop 3 at $t = 683$ days (same as in Figure~\ref{fig:Rising Loops}(d)). Color indicates amplitude of magnetic field (purple strong, yellow weak). Views are looking (a)~south along the rotation axis with grid lines in radius (in units of the solar radius) and longitude, (b)~radially inward with grid lines in longitude and latitude, (c)~westward along the axis of the magnetic wreath.}
 \label{fig:3 Views}
\end{center}
 \end{figure} 

Axial flows along loops are also seen as the loops rise through regions of faster rotation. When averaging over many loops, a net axial flow of 5.1 m s$^{-1}$ is detectable in the retrograde direction, consistent with the fluid inside the loop tending to conserve its specific angular momentum as the loop moves radially outward. Loops often become distorted as this retrograde motion interacts with the surround prograde differential rotation as the loop rise across rotational contours (see Figure \ref{fig:VrBphi}(b)).

The geometry of each loops we have examined is uniqued in its details, but Figure~\ref{fig:3 Views} shows three different perspectives on a single 3D volume-rendering of a typical loop. Loop 3, which is also shown in Figure~\ref{fig:Rising Loops}, is in the northern hemisphere and its top is roughly centered at $76^\circ$ N latitude and $12^\circ$ W longitude. Its parent wreath-segment runs slightly north-west to south-east at this location and time, causing the western foot-point to be centered further north than the eastern foot-point. The deflection away from the Equator is evident in Figures~\ref{fig:3 Views}(b and c) as the top of the loop is roughly $10^\circ$ further north than the foot-points. The roughly five-fold expansion of the loop's cross-sectional area can be seen, particularly in Figure~\ref{fig:3 Views}(c). This loop also shows an asymmetric top due to a downflow plume impacting the eastern side of the top of the loop, causing the western wide to extend further in radius.

   \begin{figure} 
 \begin{center}
 \centerline{\includegraphics[width=0.78\linewidth]{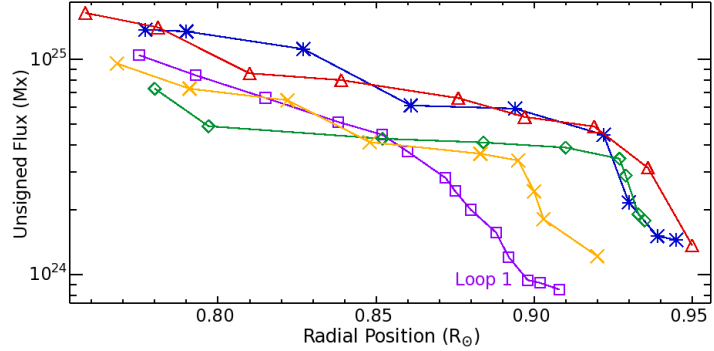}}
 \caption{Unsigned magnetic flux in five sample loops as they rise through the convective layer, including loop 1 (see Figures~\ref{fig:Rise Time} and \ref{fig:Rising Loops}).  Loops continuously loose magnetic flux through both diffusion and leakage of fluid. Their ascent is faster than the convective upflows in which they are embedded, leading to the loss of fluid and flux due to drag-like effects. Here and in the 22 loops for which detailed tracking is possible, the initial and final magnetic flux are not correlated.}
 \label{fig:Rising Flux}
\end{center}
 \end{figure} 
 
Loops start with a wide variety of field strengths and sizes and at a variety of initial radial positions.  Most loops start between 0.75 and 0.78 $R_\odot$, although loops starting as low as 0.73 $R_\odot$ are evident. When loops are traced backward in time to their starting location in order to identify the flux which will become buoyant and rise, we find that most progenitors of loops begin with about $10^{25}$ Mx of flux. The structures loose roughly 90\,\% of their flux as they rise to their peak radial positions between $0.90$ and $0.96 R_\odot$.  Much of the flux is lost as convection in a stratified fluid requires a large fraction of the fluid to overturn prior to reaching the top of the domain. Figure \ref{fig:Rising Flux} shows the magnetic flux as a function of the radial position of the top of the loop for five sample loops.  Initial flux and initial radial location do not appear to be good predictors of either final radial location or final magnetic flux.

In the specific case of loop 1, 92\,\% of the magnetic flux it started with is lost over the course of its ascent while 69\,\% of the mass flux at $0.78 R_\odot$ turns over below $0.91 R_\odot$.  The overturning mass flux carries away 61\,\% of the magnetic flux, as regions of lower field strength preferentially are lost. The next largest contributor is resistive diffusion, which dissipates 19\,\% of the initial flux. The remaining 12\,\% of the flux is lost through a combination of small-scale turbulent advection and shear. Eventually diffusive reconnection realigns the fields so that the loops are no longer distinct from the surrounding MHD turbulence.  

\subsection{Statistical Distribution of Twist and Tilt }
 
 Previous MHD simulations of flux emergence have emphasized that magnetic structures must be twisted to remain coherent as they rise \citep[see review by][]{Fan2009}.  Twist in this context can be defined by a parameter $q_A$, which for a uniformly twisted flux tube is defined as
 \begin{equation}
B_\parallel  = a_\pm \, q_A \lambda \left| \nabla \times A_\parallel \right| ,
 \end{equation}
 where $B_\parallel$ and $A_\parallel$ are, respectively, the magnetic field and magnetic vector potential along the axis of the flux tube, $a_\pm$ is 1 in the northern hemisphere and $-1$ in the southern hemisphere, and $\lambda$ is the distance from the axis of the flux tube.  For the tube to remain coherent as it rises, previous numerical simulations have suggested that twist must exceed some critical value $Q_A$ \citep{Moreno1996}.  \cite{Fan2008} used 3D simulations of buoyant magnetic structures rising through a quiescent, stratified layer and found a critical level of twist $Q_A \approx -3 \times 10^{-10} \; \mathrm{cm}^{-1}$. 

   \begin{figure} 
 \begin{center}
 \centerline{\includegraphics[width=0.5\linewidth]{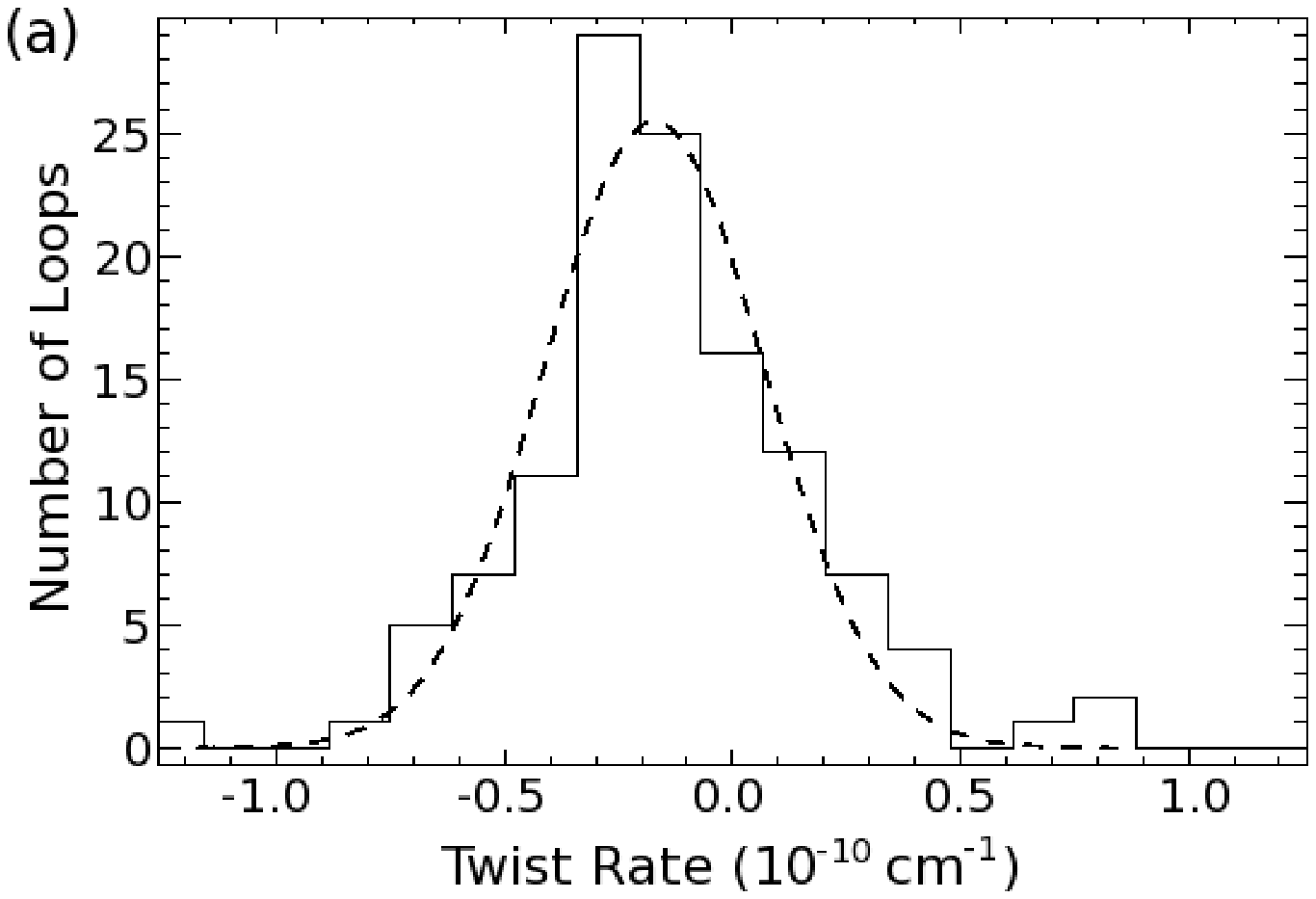}
\includegraphics[width=0.5\linewidth]{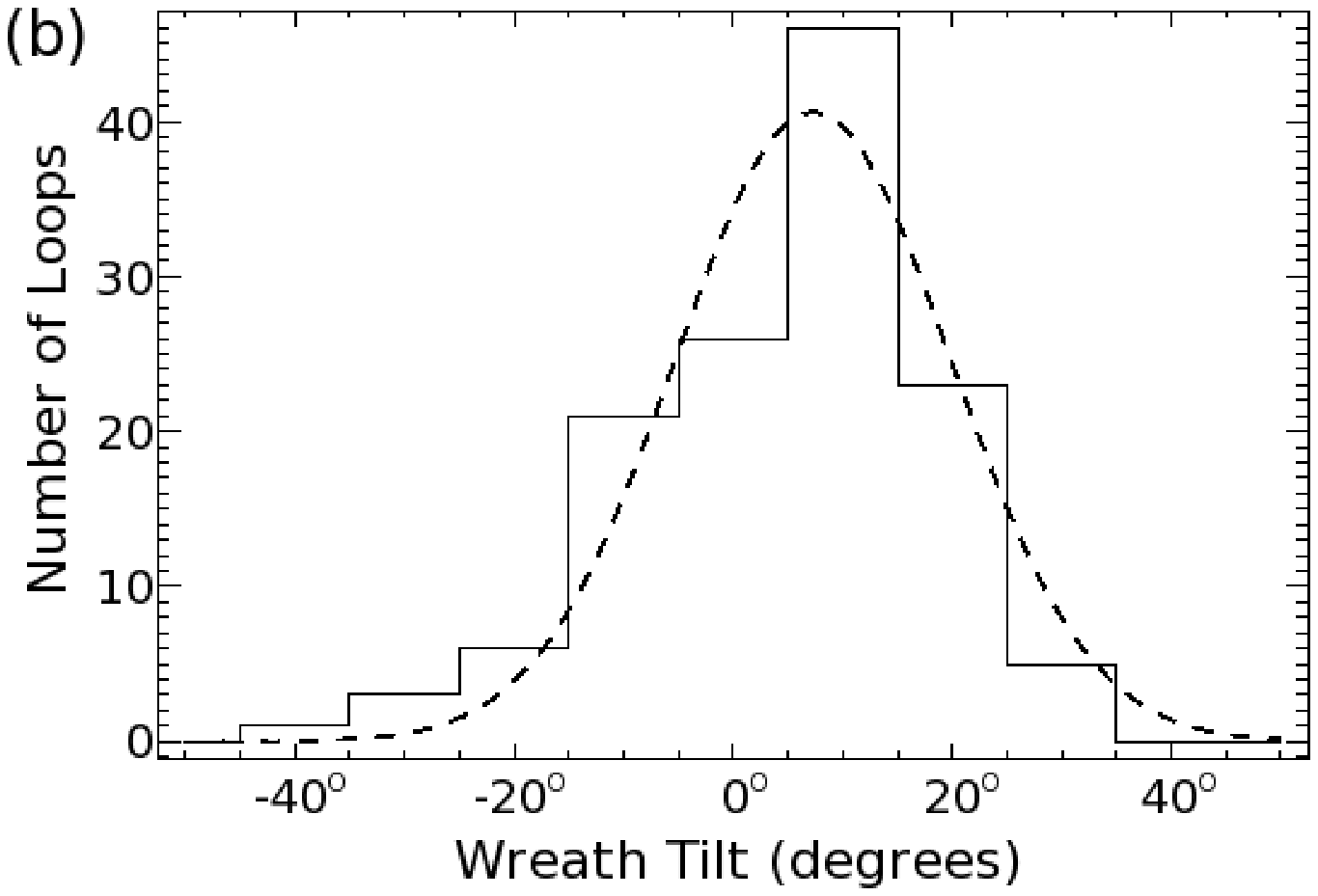} }
 \caption{(a)~Histogram of twist rate parameter $q_J$ values for the 131 loops observed in cycle 1 along with the best-fit Gaussian distribution of those values.   The distribution shows a slight preference for negative twist rates, though the mean twist rate is $(-1.8 \pm 2.4) \times 10^{-11} \; \mathrm{cm}^{-1}$. (b)~Histogram of latitudinal tilt $\Delta_\theta$ values for the same 131 loops.  Positive tilts indicate that the leading edge of the loop is closer to the Equator than the trailing edge, as used with Joy's law. Tilts have been calculated so that all values fall between $\pm 90^\circ$ for this analysis. Positive tilts are preferred and the mean latitudinal tilt is $7.3^\circ \pm 12.6^\circ$ in latitude.   }
 \label{fig:Twist Tilt}
\end{center}
 \end{figure} 

  For our simulation, the loops are clearly not uniformly twisted flux tubes, so we calculate another measure of twist following the procedure used in observational studies \citep[{\it e.g.} ][]{Pevtsov1995, Pevtsov2003, Tiwari2009}.  Sunspots often show large variations in the level and even sign of twist, so a weighted average of the twist parameter is employed, which we call $q_J$. We compute the twist parameter as 
 \begin{equation}
q_J = a_{\pm} \left[ \frac{  J_\phi }{ B_\phi } \right] ,
 \end{equation}
 where braces denote an average over radius and latitude for a longitudinal cut taken through the loop and $a_\pm$ is 1 in the northern hemisphere and $-1$ in the southern hemisphere.  We restrict our averages to contiguous regions with the correct polarity and where fields are stronger than 2.5 kG. Figure~\ref{fig:Twist Tilt}(a) shows a histogram of values for the twist parameter $q_J$ for the 131 loops identified in cycle 1, as well as the best-fit Gaussian to that distribution which peaks at $\bar{q_J} = -1.8 \times 10^{-11} \; \mathrm{cm}^{-1}$.  For comparison,\cite{Tiwari2009} report an average twist parameter of $ \bar{q_J} = -6.12 \times 10^{-11} \; \mathrm{cm}^{-1}$ for a sample of 43 sunspots. 
 
It is difficult to make a direct comparison between the two measures of twist mentioned here. In practice our loops are poorly represented by uniformly twisted tubes.  It is possible to compute the value of $q_A$ at each point in the loop and create an average value, but we find that those averages are highly sensitive to the weighting of the points and the region over which the average is taken. Alternatively, we have computed the value of $q_J$ for the formulation employed in \cite{Fan2008} and find that the value varies with the location and size of the magnetic structure in radius and latitude.  For most reasonable parameter choices, the $q_J / q_A$ is between 1 and 2. When comparing with photospheric measurements, we must also remember that considerable changes may take place as magnetic flux passes through the upper 5\,\% of the solar convection zone. The dynamics of twisted buoyant loops in that region is beginning to be studied in local domains \citep{Cheung2010}.

Of the 131 loops in cycle 1, only 13 had current-derived twist parameters $q_J$ within an order of magnitude of the critical value $Q_A$.  One explanation may be that convective upflows assisting the rise of these loops reduces the drag that they experience, thus making them less susceptible to disruption as they rise and therefore less dependent on twist for coherence. Whatever the cause, we do not see a critical value of twist beyond which loops are unable to traverse our domain.

 \begin{figure} 
 \begin{center}
 \centerline{\includegraphics[width=\linewidth]{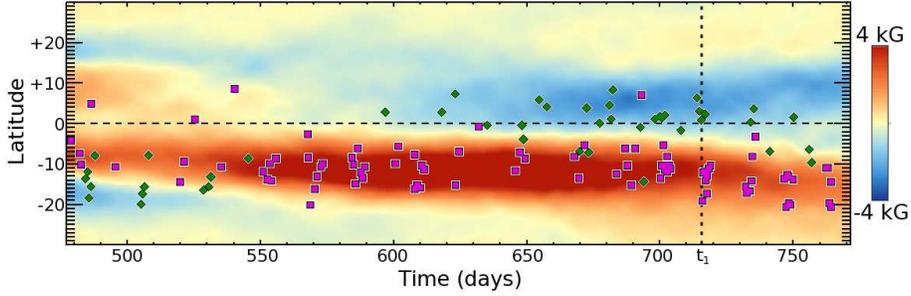}}
 \caption{Time-latitude display of toroidal magnetic field averaged in longitude and radius during the peak of cycle 1.  Over-plotted symbols indicate the time and latitude of 131 buoyant loops in the style of a synoptic map, with positive polarity loops shown as pink squares and negative polarity loops as green diamonds.  Some loops may be present from the previous cycle, particularly prior to day 550 in the southern hemisphere. Time $t_1$ at which the snapshots in Figures \ref{fig:VrBphi} and \ref{fig:Wreaths Loops} are taken is indicated by the dotted line.}
 \label{fig:LatTimeLoops}
\end{center}
 \end{figure} 

Additionally, we can look at the latitudinal tilts of the buoyant loops.  We calculate these tilts by computing the center of each loop at all longitudes where the center is within $0.02 R_\odot$ of its peak position and then fitting a linear trend to latitudinal locations of the loop center. We define positive tilts to be those with the eastern side of the loop closer to the Equator than the western side, as used in Joy's law. Here we do not consider the polarity of the loops, so values are restricted to the interval $\left[-90^\circ, 90^\circ \right]$. The distribution of tilts seen in the 131 loops found in cycle 1 is shown in Figure~\ref{fig:Twist Tilt}(b), along with the best-fit Gaussian to that distribution, which peaks at $7.3^\circ$ but is quite broad. This is similar to observations of tilts in sunspots where the trend towards Joy's law is part of broad distribution in tilt angles \citep{Li2012}.

 \section{Magnetic Cycles with Buoyant Loops}
 
Case S3 achieves three magnetic-activity cycles with reversals in global magnetic polarity.  If we define the cycle period as the time between changes in the sign of the antisymmetric components of the toroidal field at low latitudes, as in \cite{Brown2011}, then cycles 1 and 2 have periods of 278 and 269 days, respectively.  Cycle 3 had not ended at the present end of the simulation, but has been simulated for  228 days. The coexistence of cyclic magnetic activity and buoyant loops provides an opportunity to probe the relationship between axisymmetric fields which are commonly used in 2D dynamo models \citep[see review by][]{Charbonneau2010} and the buoyant transport of magnetic flux. 

We have chosen to conduct our analysis primarily using cycle 1 since the process of finding and characterizing buoyant loops is too data intensive to be carried out conveniently for all three cycles.  Figure \ref{fig:LatTimeLoops} shows a time--latitude plot of the mean toroidal field (averaged in longitude and in radius over the lower convection zone from 0.72 to 0.84 $R_\odot$), as well as the location in time and latitude of the 131 buoyant loops detected in cycle 1.  It is evident from this representation that the loops do not arise uniformly in time. Although loops tend to appear at times and latitudes when the mean toroidal fields are strong, they can also appear at times and latitudes with relatively weak mean fields. There are even examples in which loops have the opposite polarity to the longitudinally-averaged mean fields at that time and latitude.  This is consistent with the non-axisymmetric nature of the wreaths shown in Figures~\ref{fig:VrBphi}(c\,--\,d) where smaller-scale segments of intense toroidal field can be masked in the longitudinal average by larger segments of the opposite polarity.

 \begin{figure} 
 \centerline{\includegraphics[width=0.5\linewidth]{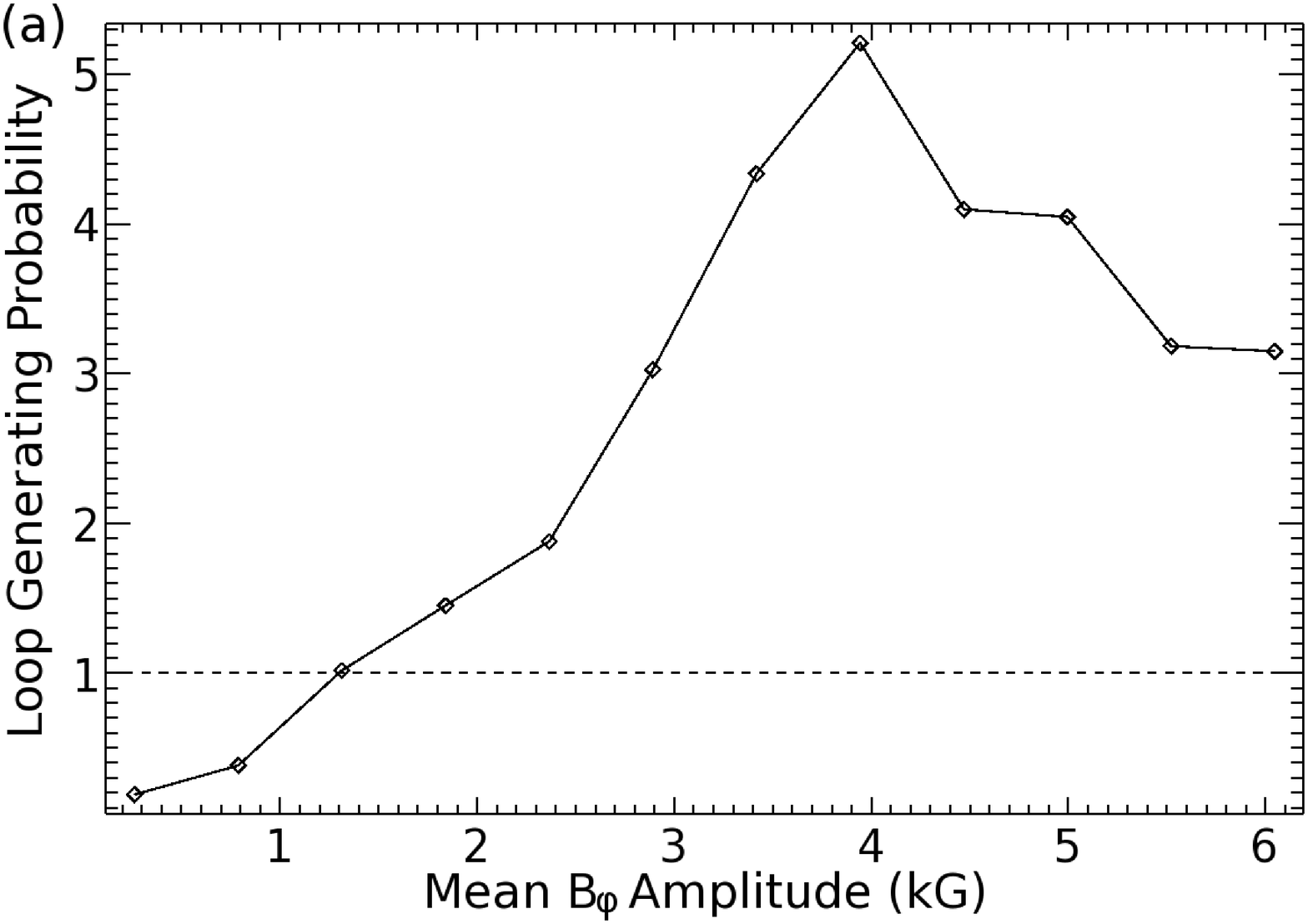}
 \includegraphics[width=0.5\linewidth]{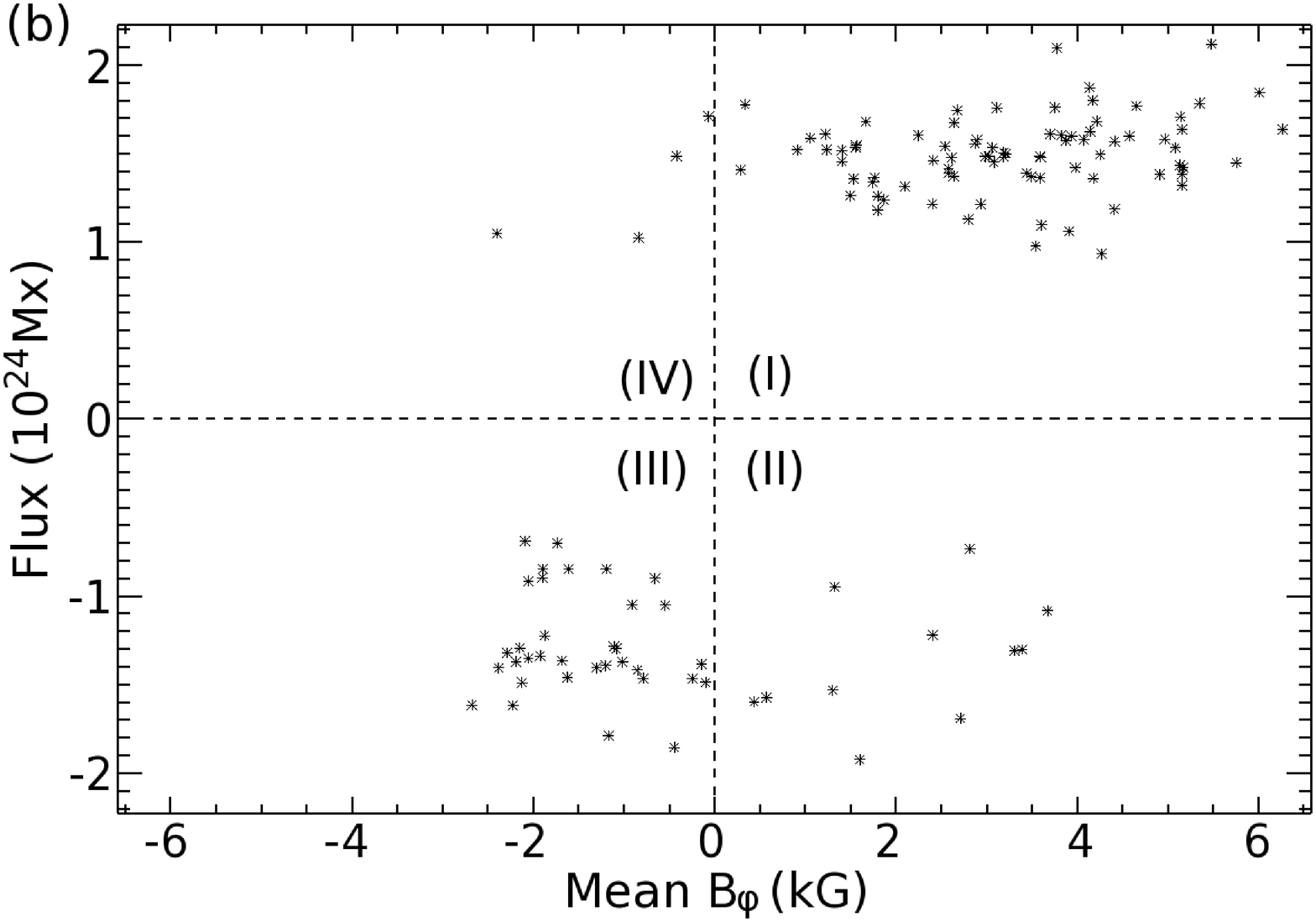}}
  \caption{(a)~Relative probability that a region with given mean $B_\phi$ will produce a buoyant loop compared to the production rate of buoyant loops in cycle 1 averaged over all events. For cycle 1 the average occurrence rate for loops was $7.6 \times 10^{-3} \; \mathrm{day}^{-1} \; \mathrm{degree}^{-1}$. We normalize all probabilities by this rate, thus the dashed line represents the average loop production rate. Note that nearly 60\,\% of the times and latitudes considered have mean field strengths of less than 1.5 kG. (b)~Total magnetic flux as a function of axisymmetric toroidal magnetic field averaged in radius at the latitude and time of each of the 131 buoyant loops in Figure~\ref{fig:LatTimeLoops}.  While most loops are associated with mean magnetic fields of the correct sense, there are 15 loops in quadrants II and IV. These loops arise from wreath segments which are canceled in the longitudinal averaging procedure by large or stronger wreath segments at the same time and latitude of the opposite polarity. }
 \label{fig:Loop Prob}
  \end{figure} 

\subsection{Relation of Loop Emergence and Mean Field Strength}
 
In many mean-field models it is assumed that buoyant magnetic flux (which can be used as a proxy for the sunspot number) at a given latitude and time is proportional to the axisymmetric toroidal field strength at that location and time at the generation depth. In particular, the Babcock--Leighton model postulates that the buoyant transport of magnetic flux occurs whenever the axisymmetric magnetic field exceeds some threshold value \citep[{\it e.g.} ][]{Durney1995, Chatterjee2004}. Here we can test this assumption by looking at the probability that a region with a given axisymmetric field strength will produce a buoyant loop. Figure \ref{fig:Loop Prob} shows the relative probability that a region with a given mean field strength will produce a buoyant loop.  Over cycle 1, the average production rate of buoyant loops is roughly one loop every two days within $30^\circ$ of the Equator.  Regions with $\langle B_\phi \rangle \le 1.5$ kG cover about 60\,\% of the time-latitude domain and produce loops at or below the average rate.  The generation probability per unit time and latitude rises to five times the mean rate for regions with $\langle B_\phi \rangle \approx 3.9$ kG.  Interestingly, the generation probability then falls for the regions of the strongest $\langle B_\phi \rangle$.  Indeed, the strongest regions of axisymmetric field are only about three times more likely to produce buoyant loops than the average production rate.  The relatively small sample size invites further study on this topic, as only 5\,\% of the domain is covered by fields above 4.2 kG, which fall in the last four bins.  However the implication that axisymmetric toroidal fields above some threshold value are less likely to produce  buoyant loops may have significant implications for mean-field models of the solar dynamo.
 
To further explore the correlation between the axisymmetric field strength and the amount of buoyant magnetic flux, we can look for correlations between the amount of flux in a given buoyant loop and the axisymmetric fields at the time and latitude of its launch. Figure \ref{fig:Loop Prob}(b) shows the magnetic flux in each of the 131 buoyant loops from cycle 1 as a function of the average value of axisymmetric toroidal field in the lower convection zone at the time of launch. Out of 131 loops, 15 were launched when the axisymmetric $B_\phi$ was of the opposite sense.  Interestingly, \cite{Stenflo2012} report that roughly 5\,\% of moderate to large active regions violate Hale's polarity law.
 
 \subsection{Preferential Longitudes for Loop Creation}
 
   \begin{figure} 
 \begin{center}
 \centerline{\includegraphics[width=\linewidth]{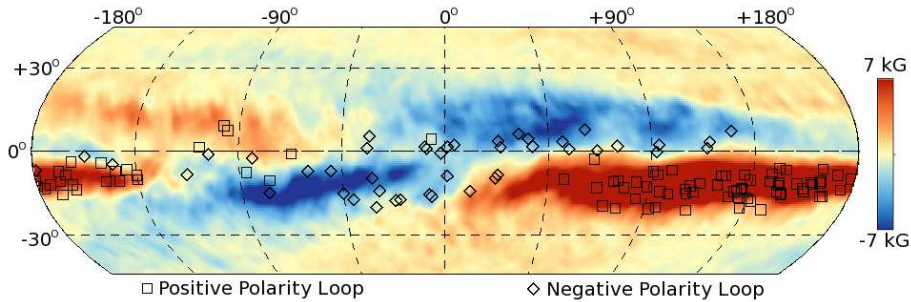}}
 \caption{Time-averaged toroidal magnetic field on a spherical surface between $\pm 45^\circ$ of latitude at $0.79 R_\odot$ during cycle 1. Symbols indicate the rotational phase in longitude of the 131 buoyant loops from Figure~\ref{fig:LatTimeLoops} at the time they were launched. Squares indicate positive polarity loops while diamonds indicate negative polarity loops. Both the wreaths and loops are confined in longitude. Loops are particularly concentrated in the strong positive wreath segment in the southern hemisphere.}
 \label{fig:Active Lons}
\end{center}
 \end{figure}
 
The longitudinal concentration of sunspots into so-called active longitudes has been observed for the past several solar cycles \citep{Henney2002}. These active longitudes provide observational evidence that the creation of buoyant magnetic structures is not a purely axisymmetric process. Magnetic wreaths in case S3 tend to be confined in longitude, as was shown in Figure~\ref{fig:VrBphi}(c).  These wreath segments are generally between $90^\circ$ and $270^\circ$ in longitude. Loops tend to be generated in these wreath segments, and thus more likely to appear in those longitudinal patches than other longitudes.  Figure \ref{fig:Active Lons} shows the time-averaged value of $B_\phi$ at $0.80 R_\odot$ over cycle 1 with the longitudinal position of the 131 buoyant loops over-plotted.  Loops are much more likely to appear over a roughly $180^\circ$ patch in longitude in the southern hemisphere. Whereas we have some longitudinal modulation, it is still far from the $10^\circ$ to $20^\circ$ confinement seen in active longitudes on the Sun.

The existence of longitudinal patches of both magnetic polarities in case S3 also provides a potential explanation for the small fraction of active regions of the ``wrong'' polarity seen in Figure~\ref{fig:Loop Prob}(b). This provides a possible mechanism for the analogous phenomena in which a small fraction of solar active regions violate Hale's polarity law. While active longitudes in the Sun are more confined than those seen here, the longitudinal confinement of the wreaths in case S3 may provide a possible pathway toward understanding active longitudes. 
  
  \section{Summary and Reflections: Buoyant Loops in Convective Dynamos}
  
 This article has explored the first global convective-dynamo simulation to achieve buoyant magnetic loops which transport coherent magnetic structures through the convection zone.  These buoyant structures arise from large-scale magnetic wreaths, which have been previously described in both persistent \citep{Brown2010} and cyclic states \citep{Brown2011, Nelson2012}.  In this work we have focused on case S3, which possesses large-scale magnetic wreaths which undergo cycles of magnetic activity and produce many buoyant magnetic loops. Case S3 was able to achieve buoyant loops due to the use of a dynamic Smagorinsky SGS model which greatly reduced diffusive processes in the simulation.
 
 Although case S3 has a rotation rate greater than the current Sun, the dynamics achieved may be applicable to solar dynamo action.  The most salient non-dimensional parameter for the creation of toroidal wreaths is the Rossby number, which considers the local vorticity $\omega$ and rotation rate $\Omega$ as Ro $= \omega / 2 \Omega$.  In case S3 the Rossby number at mid-convection zone is 0.581, indicating that the convection is rotationally constrained as is also expected in the bulk of the solar convection zone.
 
Much of the work on buoyant magnetic flux has generally regarded convection as a purely disruptive process. In our dynamo studies here, convection plays a key role in both the creation of the strong, coherent magnetic fields and the advection of magnetic flux radially outward. Turbulent intermittency provides an effective mechanism for the amplification of magnetic fields to energy densities well above equipartition with the resolved flows \citep{Nelson2012}.  Convection also assists in the transport process by the upflows helping to advect the loops. Without convection, buoyant transport of magnetic flux is generally regarded as a low-wavenumber instability on axisymmetric fields.  With convection, buoyant loops are formed on convective length scales as the result of non-axisymmetric processes. The loops realized in case S3 are not large-scale instabilities of axisymmetric flux tubes, but rather they result from turbulently amplified coherent structures becoming buoyant and being advected by convective upflows. Similar upward advection of magnetic structures by convection has been seen when considering the impact of convection on flux tubes \citep{Weber2011, Weber2012} or specified magnetic structures \citep{Jouve2009}.
  
When we consider moderate numbers of buoyant loops over an activity cycle, we find a number of trends in their collective behavior. In all of these trends, it is important to note that our statistical sample of 158 loops is significant but still relatively small. First, loops in our simulation clearly show a hemispheric polarity preference analogous to Hale's polarity law for solar active regions, although case S3 shows a slightly higher rate of violations to this trend compared to the Sun. Second, the buoyant loops tend to show latitudinal tilts similar to Joy's law for solar active regions.  As in the Sun, a wide variety of tilt angles are observed, though the average tilt angle places the leading edge of the buoyant loop closer to the Equator than the trailing edge. Third, the buoyant loops tend to show a degree of twist similar to the twist inferred from photospheric measurements of vector magnetic fields.  Again a wide variety of twist parameters are measured centered about a relatively small, negative mean value. Finally there are ranges in longitude which demonstrate repeated emergence of magnetic flux. This longitudinal modulation in the creation of magnetic loops is reminiscent of active longitudes observed in the Sun, but on larger longitudinal ranges than active longitudes in the Sun.
  
Buoyant transport of magnetic fields is a key ingredient in many models of the solar dynamo.  Mean-field models often use parameterizations to represent this buoyant transport.  We have considered connections between the axisymmetric toroidal fields in case S3 and the magnetic flux in the buoyant loops. We find that total flux in a given buoyant loop is only weakly dependent on the strength of the mean field from which that buoyant loop was generated.  Additionally, we find that the probability that a buoyant loop will be generated in regions of relatively weak mean fields is significant, and that the strongest mean fields may be less likely to generate buoyant loops than regions of moderate axisymmetric fields.

This simulation is a first step towards connecting convective-dynamo models and flux emergence in the Sun and Sun-like stars.  As we consider the role of turbulent convection, we find clear indications that it plays important roles in the dynamo that generates buoyant magnetic loops and the transport of those loops. This simulation invites continued effort towards linking convective-dynamo models and simulations of flux emergence.

%% Table
%
% \begin{table}
% \caption{}%\label{tbl:?}
% \begin{tabular}{}     
% \hline
% \multicolumn{2}{c}{<>}
% <data>
% \hline
% \end{tabular}
% \end{table}

%%%%%%%%%%%%%%%%%%%%%%%%%%%%%%%%%%%%%%%%%%%%%%%%%%%%%%%%%%%%%%%%%%%%%%%%%%%
%% Appendix
%
% \appendix   

%%%%%%%%%%%%%%%%%%%%%%%%%%%%%%%%%%%%%%%%%%%%%%%%%%%%%%%%%%%%%%%%%%%%%%%%%%%
%% Acknowledgements

 \begin{acks}
We thank Kyle Augusten, Chris Chronopolous, and Yuhong Fan for discussions and advice.  This research is partly supported by NASA through Heliophysics Theory Program grants NNX08AI57G and NNX11AJ36G. BPB is supported through NSF Astronomy and Astrophysics postdoctoral fellowship AST 09-02004.  CMSO is supported by NSF grant PHY 08-21899. MSM is also supported by NASA SR\&T grant NNH09AK14I.  NCAR is sponsored by the National Science Foundation. ASB is partly supported by both the Programmes Nationaux Soleil-Terre and Physique Stellaire of CNRS/INSU (France), and by the STARS2 grant 207430 from the European Research Council.  The simulations were carried out with NSF TeraGrid and XSEDE support of Ranger at TACC, and Kraken at NICS, and with NASA HECC support of Pleiades.  

 \end{acks}

%%% %%%%%%%%%%%%%%%%%%%%%%%%%%%%%%%%%%%%%%%%%%%%%%%%%%%%%%%%%%%
%% Bibliography
%
% Using BibTeX
%
% \bibliographystyle{spr-mp-sola}
 \bibliographystyle{spr-mp-sola-cnd} %% Alternative style: no title, no concluding page
 \bibliography{SoPhys_Loops}  
%
% Without BibTeX 
% \begin{thebibliography}{}
% \bibitem[\protect\citeauthoryear{Author}{Year}]{key}
%   <bibliographical entry>
%
% \bibitem[\protect\citeauthoryear{}{}]{}
%   
%  
% \end{thebibliography}

\end{article} 
\end{document}